# Libros en abierto de las editoriales universitarias españolas

## Open books from Spanish university presses

**Rosana López-Carreño; Ángel M. Delgado-Vázquez; Francisco-Javier Martínez-Méndez**

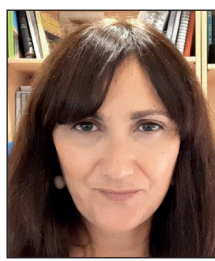

**Rosana López-Carreño** ✉
*https://orcid.org/0000-0002-2097-9389*

*Universidad de Murcia
Fac. de Comunicación y Documentación
Departamento de Información y Documentación
30100 Espinardo (Murcia), España
rosanalc@um.es*

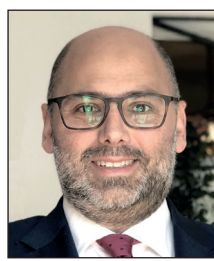

**Ángel M. Delgado-Vázquez**
*https://orcid.org/0000-0003-2461-8553*

*Universidad Pablo Olavide
Área de Biblioteconomía y Documentación
Ctra. de Utrera, Km 1
41013 Sevilla, España
adelvaz@upo.es*

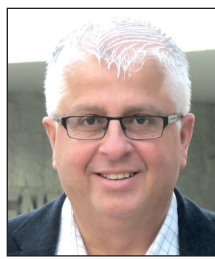

**Francisco-Javier Martínez-Méndez**
*https://orcid.org/0000-0003-1098-9361*

*Universidad de Murcia
Facultad de Comunicación y Documentación
Departamento de Información y Documentación
30100 Espinardo (Murcia), España
javima@um.es*

**Resumen**

Este trabajo analiza el conjunto de las publicaciones científicas en acceso abierto, distintas de las revistas científicas (monografías, actas de congresos, materiales didácticos y literatura gris), dispuestas para su consulta por las editoriales universitarias públicas, estudiando su volumen, tipología documental, nivel de descripción y políticas de acceso abierto con el objetivo de medir el grado de incorporación y cumplimiento de los principios de Ciencia Abierta. Se ha llevado a cabo una exhaustiva revisión del material publicado en acceso abierto por estas editoriales que ha permitido establecer un diagnóstico de su nivel de edición en acceso abierto. La literatura gris es el tipo documental más frecuente seguido de la monografía, en la publicación en abierto de las editoriales universitarias que no alcanza ni el 5% de la producción editorial universitaria. Los resultados permiten concluir que la publicación académica, y más concretamente el libro en acceso abierto, sigue teniendo una presencia muy reducida dentro de la producción editorial de estas instituciones.

**Palabras clave**

Publicaciones en acceso abierto; Ciencia abierta; Editoriales universitarias; Publicación electrónica; Publicación digital; Libros en abierto; Libros académicos; Monografías; Indicadores de calidad; Publicaciones académicas; Políticas.

**Abstract**

This paper analyses the set of scientific publications in open access, other than journals (monographs, conferences proceedings, teaching materials and grey literature), published by Spanish public universities, studying their volume, documentary typology, level of description and open access policies with the aim of measuring their degree of incorporation and compliance with the principles of Open Science. An exhaustive review of the disposed material in open access by these publishers has been carried out, which has allowed to make a diagnosis of their level of open access publishing. Grey literature is the most common documentary type followed by the monograph, in the open publication of these publishers that does not reach even 5% of the average editorial production. The results allow us to conclude that the academic publishing, and more specifically the academic books in open access, still has a very reduced presence within the editorial production of these institutions.

## 1. Introducción

El movimiento por el acceso abierto de la información científica es uno de los procesos más influyentes en la comunicación científica desde principios de este siglo, habiendo propiciado significativas modificaciones en la difusión y acceso a los resultados de investigación (**Abadal**, 2012). En España, el primer paso fue la apuesta de la Administración Central por el movimiento de acceso abierto sustanciada en la *Ley 14/2011* (*España*, 2011), *de 1 de junio, de la ciencia, la tecnología y la innovación* (*Fecyt*, 2016; *Rebiun*, 2019) que pretendía impulsar, por parte de las entidades de I+D+i, los repositorios de acceso abierto de sus publicaciones científicas y el establecimiento de sistemas que los conectaran con iniciativas similares (*España*, 2011). En Europa, esta apuesta tiene su reflejo en el mandato inserto en el *Programa marco de financiación de la ciencia Horizonte 2020* para que todo beneficiario garantice el acceso abierto a todas las publicaciones científicas revisadas por pares relacionadas con sus resultados de investigación, asegurándose de que cualquier publicación científica revisada por pares pueda leerse, descargarse e imprimirse en línea. Si bien el tipo documental mayoritario es el artículo de revista también se recomienda brindar acceso abierto a otras publicaciones científicas como monografías, libros, actas de congreso, informes, etc. (*European Commission*, 2017, p. 5) sin olvidar el depósito de los datos de investigación en repositorios institucionales, temáticos o centralizados bajo licencias abiertas del tipo Creative Commons (preferentemente CC BY o CC0).

Estas iniciativas se han materializado en dos líneas de trabajo: el depósito y la publicación. El depósito en repositorio es la "vía verde" del acceso abierto. La otra línea, la "vía dorada", persigue la publicación de los artículos en revistas de acceso abierto (**Chan** *et al.*, 2001). Este modelo aumenta de forma paulatina su importancia dentro del sector editorial, si bien ha experimentado cierta diversificación, dando lugar a otras vías y denominaciones como "diamante", "híbrida", "bronce", incluso "negro" (**Björk**, 2017; **Martín-Martín** *et al.*, 2018) en las que, en muchos casos, los autores asumen parte de los costes de la publicación.

En las editoriales universitarias se debate sobre la circulación y la venta del libro académico. Los usos editoriales tradicionales en la distribución siguen vigentes, pero se están explorando otras vías de aproximación al público lector, abriendo

> "nuevos espacios para la interlocución de saberes, uno de esos espacios es por ejemplo [...] el acceso abierto, el cual puede constituirse en un amplificador de la labor editorial" (**Córdoba-Restrepo** *et al.*, 2018, p. 5).

Cordón-García ya planteaba que

> "la monografía científica no puede apartarse del flujo global de la comunicación académica, sino que ha de estar fuertemente imbricada en la misma. El libro electrónico constituye una oportunidad única para favorecer este encuentro" (**Cordón-García**, 2014, p. 270).

En las universidades españolas estas iniciativas se han implementado por medio de dos vías paralelas, independientes entre sí:

- a través de los repositorios institucionales, gestionados por las bibliotecas universitarias, se facilita el depósito, el autoarchivo ("vía verde") e incluso, aunque de manera minoritaria, la publicación de las obras generadas en su institución;
- las editoriales universitarias han iniciado el proceso de apertura de los libros académicos, propiciando la publicación de resultados de investigación a través de la "vía dorada". Estos libros, según Urbano,

    > "responden al perfil de la monografía académica, entendida como vehículo de comunicación de resultados de investigación, y para la que el acceso abierto se puede ver como una extensión lógica de su exitosa aplicación a las revistas científicas" (**Urbano**, 2018, p. 30).

Las bibliotecas y editoriales universitarias deben tener definidas políticas, requisitos y procedimientos sobre el acceso abierto de su producción científica, dando visibilidad a las obras tanto depositándolas en sus repositorios institucionales como publicándolas en la propia web de la editorial. Para ello es preciso:

- Establecer políticas de acceso abierto normalizadas para sus repositorios y servicios editoriales, así como la necesaria adopción unívoca del uso de licencias abiertas, preferentemente Creative Commons (CC), para los documentos a archivar y a publicar.
- Identificar de forma clara los tipos documentales a publicar en abierto que requieran de revisión por pares (ejemplo: monografías o informes), así como aquellos que por su naturaleza (premios, homenajes, etc.) no requieran de dicha revisión.
- Combinar la publicación con los modelos de negocio de las editoriales universitarias (por ejemplo, suscribir con los autores contratos de embargo temporales y proceder a la publicación en abierto transcurrido un tiempo). De esta forma se aumenta el conjunto de publicaciones potenciales a difundir por esta vía.

- Definir el flujo editorial y de apertura de las publicaciones en abierto por el transcurrir de las obras para que el autor/investigador obtenga las garantías de reconocimiento de su publicación.
- Imponer la obligatoriedad de la utilización de identificadores persistentes y unívocos, como orcid, DOI o handle, renunciando al uso del código ISBN como identificador principal ya que su finalidad es el control comercial de la monografía y esto carece de sentido en el acceso abierto. Con el uso de orcid se mejora el control de autoridades, así como

    "su integración e interoperabilidad en plataformas bibliográficas, gestores editoriales y otros perfiles de autores e investigadores" (**Martínez-Méndez**; **López-Carreño**, 2019, p. 91)

    y, con el DOI, se intenta facilitar la accesibilidad y la extracción de referencias de las publicaciones en abierto.
- Usar esquemas de metadatos, como *Dublin Core* o *DataCite*, para lograr la interoperabilidad entre repositorios y sistemas de gestión editorial con la idea de aumentar la disponibilidad de metadatos abiertos en las publicaciones académicas:

    "*Crossref*, *I4OC* y *OpenCitations* han jugado un papel crucial en estos desarrollos. Han centrado su atención principalmente en hacer datos de citas disponibles abiertamente" (**Waltman**, 2019).
- Extraer y visibilizar las referencias incluidas en estas publicaciones, de manera similar a como se hace con las revistas científicas, para ser más fácilmente rastreadas y contabilizadas por los motores de búsqueda y las plataformas bibliográficas:

    "Una comparación de *Crossref* con *Web of Science* y *Scopus* mostró que faltan decenas de millones de referencias en *Crossref* porque los editores no pudieron depositarlas" (**Van-Eck** *et al.*, 2018).
- Identificar las entidades y proyectos de investigación financiados ligados a los libros en abierto. Esta información deberá recogerse en un campo propio de su descripción (al igual que se hace con los artículos de las revistas) para evidenciar la rentabilidad de la producción científica subvencionada.
- Registrar y agregar contenidos a otros repositorios y contenedores de publicaciones en acceso abierto temáticos de monografías, como *Oapen Library* o *Directory of Open Access Books* (*DOAB*), para la obtención de una mayor visibilidad.
- Implementar, de forma unificada, métricas de uso en los sistemas de gestión de las publicaciones abiertas sincronizadas con las propias de los repositorios.

Las monografías publicadas por las editoriales universitarias se están revalorizando gracias al interés que han despertado en las agencias nacionales de evaluación. *Aneca* puso en marcha hace poco tiempo el sello o mención de calidad editorial CEA-APQ con el objetivo de

"reconocer las mejores prácticas dentro de la edición universitaria española y convertirse en un signo distintivo que tanto las agencias de evaluación de la actividad investigadora como la comunidad académica e investigadora podrán identificar fácilmente" (*UNE*, 2020).

Este sello de calidad es exclusivo para la valoración de colecciones de las editoriales universitarias, servicios que, hasta ahora, han estado más centrados en la publicación impresa comercial incentivando apenas un poco la publicación en acceso abierto. Teniendo en cuenta que estas editoriales

"constituyen el primer grupo de edición académica de España, por delante de otros como *Hachette* o *Planeta*, ya que en 2015 publicaron el 30% de los libros académicos" (**Abadal**; **Ollé**; **Redondo**, 2018, p. 301) es imprescindible establecer estrategias coordinadas entre las universidades en torno al libro académico en abierto como medio de difusión de la actividad científica. A partir de la afirmación anterior, si el volumen de publicación de monografías en acceso abierto es bajo en las editoriales universitarias, lo normal es que lo sea en el conjunto de todas las editoriales españolas. Si las editoriales más cercanas a los investigadores no fomentan las monografías en abierto va a ser complicado que lo haga una editorial comercial.

**Abadal**, **Ollé** y **Redondo** en el trabajo anterior (2018, p. 308), concluyen que, si bien

"el número de editoriales que publican en acceso abierto es notable, el número de títulos no lo es, y no se dispone de plataformas que permitan un acceso integrado y una buena visibilidad de la oferta".

En el ámbito latinoamericano, **Córdoba-Restrepo** *et al.* (2018) consideran poco consolidada la edición en abierto de libros académicos, sólo 31 editoriales universitarias (el 21%) aportaban alguna información sobre su modelo de negocio o la protección de las obras, es decir, con las licencias para su uso, copia, difusión, etc. Ahondando en esto, el informe elaborado por el grupo *E-Lectra* sobre las Universidades asociadas a la *Unión de Editoriales Universitarias Españolas* apunta a que el porcentaje de aquellas que han publicado algún libro en acceso abierto es todavía bajo, el 57,1% (**Cordón-García**, 2019).

La disposición en abierto de las monografías académicas es un hecho que tendrá que replantearse a corto plazo como ya ocurre en otros países, como por ejemplo en Reino Unido cuyo

"organismo de financiación para la investigación e innovación (*UKRI*) propone extender un requisito de acceso abierto a las monografías académicas y capítulos de libros a partir de enero de 2024" (**Page**, 2020);

o en EUA, con la iniciativa *TOME* cuyo objetivo es

"apoyar la publicación digital de libros académicos revisados por pares por las editoriales universitarias participantes" (**Ruttenberg**, 2019).

En el caso de Alemania, el proyecto *Open-Access-Hochschulverlag* de la *Leipzig University of Applied Sciences*, establece un minucioso flujo de trabajo para la publicación de monografías de acceso abierto buscando su rentabilidad y eficiencia (**Schrader** *et al.*, 2020), que podría adaptarse en las editoriales universitarias españolas. El estudio citado del grupo *E-Lectra* recoge la opinión del 80% de los editores universitarios, que expresan su preocupación por la pervivencia del libro académico, y lo achacan sobre todo a:

- sistemas de evaluación científica;
- bajada en el número de lectores;
- superior uso de las revistas como medio de comunicación de los resultados de investigación.

Poco más de un tercio de estos servicios ven en el libro digital en acceso abierto una tabla de salvación (**Cordón-García**, 2019).

## 2. Definición del libro académico abierto

La *Ley 23/2011, de 29 de julio, de depósito legal*, define libro como la

"obra científica, artística, literaria o de cualquier otra índole que constituye una publicación unitaria en uno o varios volúmenes y que puede aparecer impresa o en cualquier soporte susceptible de lectura".

El texto legal considera libros (como no podía ser de otro modo) a los electrónicos y los que se publiquen o se difundan por internet. Esta definición genérica no parece ser suficiente en el entorno de la ciencia abierta y lo que se espera del libro académico abierto, que debería estar alineado con la revista académica en abierto para poder ser considerado como una monografía académico-científica, resultado de una investigación muchas veces financiada y sometida a criterios de garantía de calidad editorial (incluyendo, aunque no sólo, la revisión por pares). Estos libros se ponen a disposición de la comunidad científica en repositorios u otros contenedores para su uso libre y gratuito.

Para cuantificar el volumen de la publicación de estas monografías en abierto y hacernos una idea de su importancia, disponemos de *Dimensions*, base de datos con amplia cobertura que incluye las monografías editadas en colaboración, y que permite filtrar por publicaciones en acceso abierto (**Orduña-Malea**; **Delgado-López-Cózar**, 2018; **Thelwall**, 2018; **Harzing**, 2019).

Tabla 1. Datos totales de monografías y libros editados en colaboración

|  | **N. de monografías** | **N. en OA** | **Depositadas** | **Vía verde, preprints y postprints** | **Publicadas** |
|---|---|---|---|---|---|
| Monografías | 778.766 | 242.680 (31,16%) | 16.288 | 148.521 | 77.871 |
| Libros editados (en colaboración) | 292.199 | 37.754 (12,92%) | 6.294 | 18.588 | 12.917 |

El gráfico 1 muestra la evolución de las monografías y libros publicados desde el año 2010 con datos procedentes de *Dimensions*, así como la cifra anual de este tipo de documentos disponibles en acceso abierto.

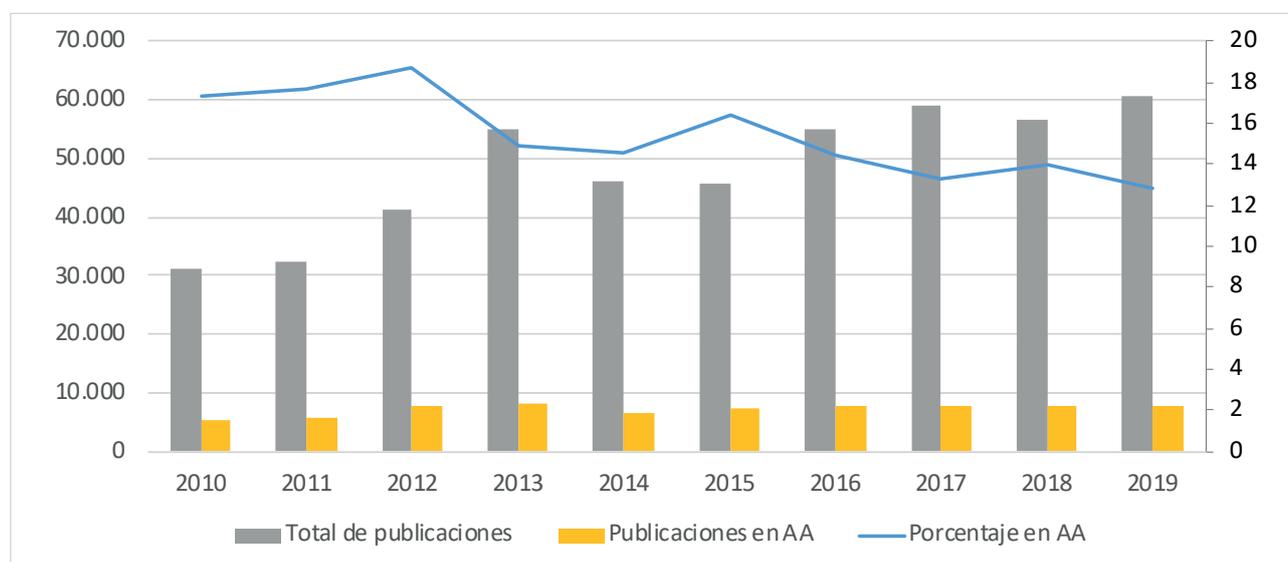

Gráfico 1. Evolución de libros (monografías y libros editados) totales y en acceso abierto, en el mundo, según *Dimensions* (2010-2019)

Como se aprecia, el porcentaje es relativamente bajo (no llega al 15%), y no ha experimentado variaciones notables a lo largo de la década (incluso en 2019 descendió ligeramente). En el caso de España no se disponen de datos precisos de la publicación en abierto de los libros de las universidades públicas, más allá del porcentaje que sobre el total de libros académicos publicados aportan **Abadal**, **Ollé** y **Redondo** (2018).

Paliar esta deficiencia de datos es uno de los objetivos que persigue el presente trabajo, además de observar el grado de disposición de libros y monografías en acceso abierto en las colecciones editadas por las editoriales universitarias públicas, estudiando su volumen, tipología documental, descripción y políticas de acceso abierto para conocer el grado de incorporación y cumplimiento de los preceptos de ciencia abierta.

### 3. Metodología

El primer paso consistió en visitar las sedes web de las universidades públicas españolas y localizar las webs de sus respectivas editoriales. En esta revisión se observó que 45 universidades disponen de editorial propia y únicamente 20 de ellas publican libros y monografías en abierto, agrupadas en un mismo espacio o permitiendo su localización a través de un buscador interno en la web de la editorial, constituyendo este subgrupo nuestro objeto de estudio: *Autónoma de Madrid*, *A Coruña*; *Cantabria*, *Castilla-La Mancha*, *Córdoba*, *Extremadura*, *Girona*, *Granada*, *La Rioja*, *Málaga*, *Murcia*, *Salamanca*, *Santiago de Compostela*, *Valencia*, *Valladolid*, *País Vasco*, *Internacional de Andalucía*, *Jaume I*, *Politècnica de València* y *Rovira i Virgili*.

El segundo paso fue recopilar información relativa a:

- uso de identificadores en las publicaciones en abierto: DOI, orcid e ISBN;
- modo de alojamiento y uso del texto completo, anotando si es vía descarga directa de la web de la editorial, repositorio institucional o por medio de un gestor de contenidos como, por ejemplo, *OMP* (*Open Monograph Press*), el más difundido, ya que es de *PKP*, la misma institución canadiense que produce *OJS*, el gestor de revistas científicas;
- política de acceso abierto: si se indican los requisitos y procedimientos en torno al acceso abierto de sus obras;
- tipo de licencia Creative Commons utilizada: CC BY/ CC BY-SA/ CC BY-ND/ CC BY-NC-ND. Así se dispondrá de información sobre el tipo de licencia más frecuentemente empleada en las editoriales universitarias.

El último paso consistió en comprobar las publicaciones en acceso abierto en las editoriales objeto de estudio. Esta fase fue laboriosa porque ninguna ofrecía esta distinción y ha habido que visualizar uno a uno los documentos publicados en la web. En este proceso se descartó a la *Universidad de Salamanca* por resultar imposible este filtrado y el posterior cómputo. En este último paso, se descartaron las revistas porque nuestro interés radicaba en los libros y monografías, principalmente, junto con las actas de congresos, material didáctico y literatura gris (informes, exposiciones, compilaciones, homenajes, lecciones de apertura, premios, tesis o mapas).

La observación y recolección de datos se llevó a cabo durante el mes de febrero de 2020.

### 4. Resultados

Respecto al uso de identificadores persistentes destaca el bajo nivel de uso de identificadores de autores (orcid sólo es utilizado por el 20% de las editoriales), así como del registro de DOI (sólo un 25% lo consigna). Estos valores están muy lejos de las revistas científicas en abierto donde el uso del DOI es generalizado y orcid es una tendencia emergente (**Martínez-Méndez**; **López-Carreño**, 2019). Es prácticamente unánime la asignación del ISBN, con la excepción de la editorial de la *Universidad de Castilla-La Mancha* que no lo asigna a sus publicaciones en abierto y opta por el DOI para identificar sus publicaciones. Esto se debe a la digitalización de publicaciones impresas o a los usos y costumbres de la gestión editorial tradicional en papel, así como a las demandas de los autores en el requerimiento del ISBN para su publicación a efectos meritorios.

Diez de las editoriales universitarias publican directamente en la web. Sólo tres de ellas (*Autónoma de Madrid*, *Murcia* y *Rovira i Virgili*) lo hacen por medio de un sistema gestor de contenidos: *Open Monograph Press* (*OMP*) de *PKP*. Este escaso nivel de uso está motivado por la lenta evolución de esta aplicación de código abierto como, por ejemplo, la necesidad de incorporar en su gestión la fase de revisión de los manuscritos (**Fruin**, 2019). El 30% de las editoriales (5 en total) deposita sus publicaciones en abierto en sus repositorios institucionales, fórmula alternativa válida siempre que dichos repositorios cumplan unos mínimos requisitos en cuanto al uso de identificadores (URI, orcid o DOI/Handle) y garanticen una básica descripción de metadatos.

En ningún caso se lleva a cabo la extracción de las referencias bibliográficas de las propias publicaciones en abierto por parte de las editoriales, algo habitual en las revistas. Esto se debe, fundamentalmente, a no disponer de un sistema de gestión que lo permita, debidamente implementado en el proceso editorial. Aunque existen soluciones de software libre para la gestión editorial de monografías, como el gestor *OMP* antes citado, este aspecto (extracción de la bibliografía) no se ha considerado todavía en la descripción de este tipo de publicación, decisión contraproducente porque es esencial para el recuento de las citas y la visibilidad de las referencias bibliográficas. Llama poderosamente la atención que tampoco se realiza esto a nivel de los repositorios institucionales, sistemas donde sí es posible extraer las referencias citadas para posibilitar su posterior rastreo y agregación por parte de buscadores y contenedores, algo fundamental en el paradigma de intercambio de información bibliográfica.

Tabla 2. Uso de identificadores en las publicaciones en abierto de las editoriales universitarias

| Universidades públicas | URL editorial | DOI | Orcid | ISBN |
|---|---|---|---|---|
| Autónoma de Madrid (UAM) | https://libros.uam.es | No | No | Sí |
| A Coruña (UDC) | https://www.udc.es/es/publicacions | Sí | No | Sí |
| Cantabria (Unican) | https://www.editorial.unican.es | Sí | No | Sí |
| Castilla-La Mancha (UCLM) | http://publicaciones.uclm.es | Sí | No | No |
| Córdoba (UCO) | https://www.uco.es/ucopress | No | No | Sí |
| Extremadura (UEX) | https://www.unex.es/organizacion/servicios-universitarios/servicios/servicio_publicaciones | No | No | Sí |
| Girona (UDG) | http://www.udg.edu/es/viu/serveis-universitaris/llibres-i-revistes-udg | No | No | Sí |
| Granada (UGR) | https://editorial.ugr.es | No | No | Sí |
| La Rioja (UR) | https://publicaciones.unirioja.es | No | No | Sí |
| Málaga (UMA) | https://www.umaeditorial.uma.es | No | Sí | Sí |
| Murcia (UM) | https://www.um.es/web/editum | No | No | Sí |
| Salamanca (USAL) | https://eusal.es/index.php/eusal | No | No | Sí |
| Santiago de Compostela (USC) | http://www.usc.es/es/servizos/publicacions | No | No | Sí |
| València (UV) | https://puv.uv.es | No | No | Sí |
| Valladolid (UVA) | http://www.publicaciones.uva.es | No | Sí | Sí |
| País Vasco (UPV/EHU) | https://www.ehu.eus/es/web/argitalpen-zerbitzua/home | No | No | Sí |
| Internacional de Andalucía (UIA) | https://www.unia.es/publicaciones | No | No | Sí |
| Jaume I (UJI) | https://www.uji.es/serveis/scp | Sí | Sí | Sí |
| Politècnica de València (UPV) | http://www.upv.es/entidades/AEUPV | No | No | Sí |
| Rovira i Virgili (URV) | http://www.publicacions.urv.cat | Sí | No | Sí |
| Subtotales (Sí) | | 5/20 (25%) | 4/20 (20%) | 19/20 (95%) |

A pesar de la disposición de las publicaciones en abierto, sólo el 30% de las editoriales que forman parte del estudio hace alguna mención a su política de acceso abierto, requisito obligatorio en las revistas y, de manera progresiva, una exigencia en directorios de libros como *DOAB* o la plataforma *Oapen*. En este sentido, destacan las universidades *Politècnica de València* y *Rovira i Virgili* por su clara apuesta hacia el acceso abierto dentro de sus políticas institucionales (además de la puesta en marcha del repositorio institucional y del portal de revistas científicas, estas universidades han aprobado mandatos para el acceso abierto, se han incorporado a manifiestos internacionales, han llevado a cabo campañas de sensibilización y de ayuda para el fomento de la publicación en acceso abierto, entre otras actividades que demuestran su interés por este movimiento). El bajo porcentaje identificado entre las editoriales de las universidades españolas entra en claro conflicto con las declaraciones de apoyo al acceso abierto a la producción científica que sus consejos de gobierno aprobaron hace cuatro o cinco años.

Por otra parte, destaca negativamente que algo más de un tercio de las editoriales analizadas no indique la modalidad de licencia Creative Commons asignada a las publicaciones en abierto. Este hecho llama poderosamente la atención porque estas publicaciones no van a ser objeto de venta y se debería indicar a los lectores la posibilidad de uso y reutilización de sus contenidos. Entre las que sí

Tabla 3. Políticas y licencias de acceso abierto establecidas en las publicaciones en abierto por las editoriales universitarias

| Editorial | Política OA | Licencia *Creative Commons* |
|---|---|---|
| UAM | Sí | CC BY-NC-ND |
| UDC | No | CC BY NC ND |
| Unican | No | No |
| UCLM | Sí | CC BY-NC-ND |
| UCO | No | No |
| UEX | Sí | CC BY-NC-ND |
| UDG | No | CC BY-NC-ND |
| UGR | No | No |
| UR | Sí | CC BY-NC-ND |
| UMA | No | CC BY-NC-ND |
| UM | No | No |
| USAL | No | CC BY-NC-ND |
| USC | No | CC BY-NC-ND |
| UV | No | No |
| UVA | No | CC BY-NC-ND |
| UPV/EHU | No | No |
| UIA | No | No |
| UJI | No | CC BY SA |
| UPV | Sí | CC BY |
| URV | Sí | CC BY-NC-ND |
| Subtotales<br>Sí (6/20) = 30%<br>No (14/20) = 70% | | CC BY-NC-ND 60%<br>CC BY SA 5%<br>CC BY 5%<br>No 35% |

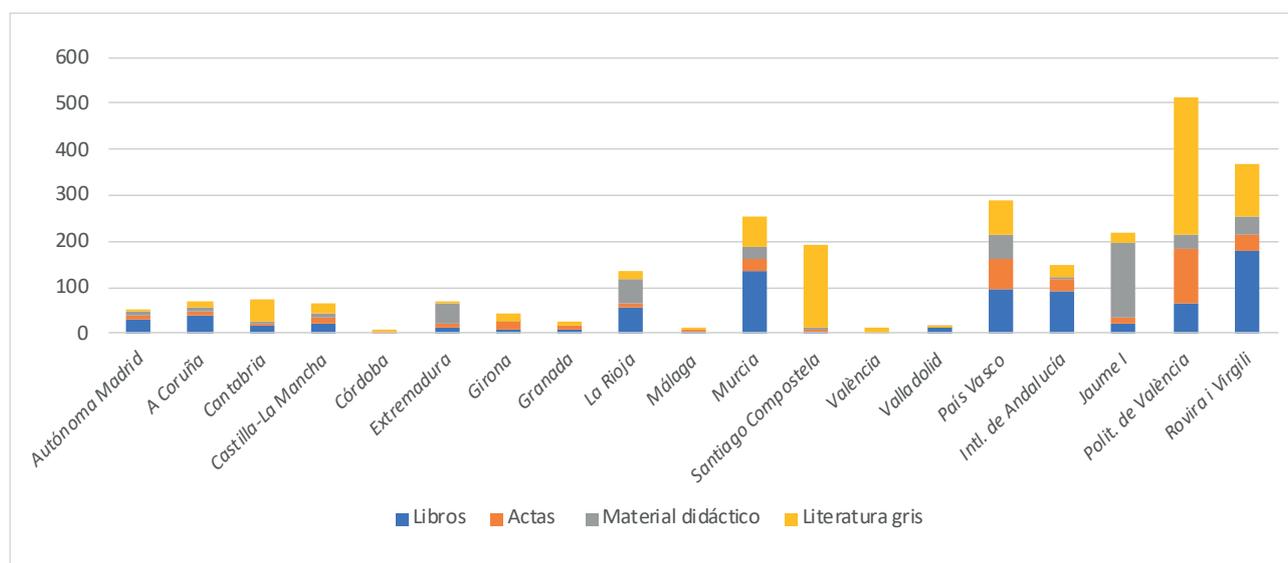

Gráfico 2. Distribución de número de publicaciones en abierto en total por tipo documental hasta febrero de 2020

informan, es mayoritaria (en un 60% delos casos) la modalidad CC BY-NC-ND (atribución, no uso comercial y no obra derivada). Esto tampoco es positivo porque esta modalidad es restrictiva respecto a su uso y va en contra de la línea futura trazada en el *Plan S* (**Hernández-Pérez**, 2019). La excepción es la *Universitat Politècnica de València* (*UPV*) que utiliza la licencia más abierta de las disponibles (BY: atribución).

En cuanto a la producción editorial, tal como se observa en el gráfico 2, la *UPV* es la que tiene más publicaciones en acceso abierto (515), seguida de la *Rovira i Virgili* (370), *País Vasco* (288) y *Murcia* (252). Estas cifras, sin ser muy altas, contrastan enormemente con el escaso volumen de las editoriales de *Valladolid* (16), *València* (14), *Málaga* (10) y *Córdoba* (9). La *UPV* es además la única de entre las cinco politécnicas españolas que edita monografías en acceso abierto.

Respecto a los tipos documentales predominantes publicados en acceso abierto en las editoriales analizadas, el más frecuente es la literatura gris (936) seguido de los libros (805), material didáctico (443) y actas de congresos (388). Es revelador que los libros no sean el tipo predominante entre las publicaciones en abierto, cuando, junto con las revistas científicas, son primordiales y primigenios en las editoriales universitarias. Esto induce a pensar que el acceso abierto ha servido para dar salida, además de a trabajos académicos de alto nivel, a otro tipo de monografías de complicada distribución comercial y cuyas ediciones no siempre han formado parte de una política editorial clara del servicio. También han tenido hueco en los portales de acceso abierto obras fruto de compromisos de las autoridades académicas universitarias y cuya publicación no suele estar ligada a proceso alguno de revisión de su calidad. Esta tesis se refuerza con el hecho de que la literatura gris sea el tipo documental con mayor número de publicaciones en acceso abierto, síntoma de que esta vía de acceso a la publicación ha sido, hasta ahora, una opción para la producción científica que quedaba fuera del proceso editorial tradicional. Las editoriales universitarias que destacan por la publicación de libros en abierto son, en primer lugar, la *Rovira i Virgili* (178 libros), seguida de *Murcia* (134), *País Vasco* (95) e *Internacional de Andalucía* (91). En el otro extremo, las universidades de *Córdoba* y *Valencia* no disponen de ningún libro en abierto en la web de sus editoriales (con independencia del uso del repositorio institucional).

La tabla 4 muestra el porcentaje de títulos publicados en abierto frente al total de los editados por las universidades públicas españolas. Con la excepción de las universidades de *La Rioja* y *Jaume I* (y quizá, pero con

Tabla 4. Número de registros (ISBNs) totales de cada editorial universitaria

| Editorial | ISBNs totales | Abiertos | % abierto |
|---|---|---|---|
| *UAM* | 1.633 | 53 | 3,25 |
| *UDC* | 4.652 | 70 | 1,50 |
| *Unican* | 1.059 | 74 | 6,99 |
| *UCO* | 1.448 | 9 | 0,62 |
| *UEX* | 1.122 | 72 | 6,42 |
| *UDG* | 754 | 42 | 5,57 |
| *UGR* | 7.722 | 24 | 0,31 |
| *UR* | 296 | 136 | 45,95 |
| *UMA* | 2.104 | 10 | 0,48 |
| *UM* | 2.577 | 252 | 9,78 |
| *USC* | 2.173 | 194 | 8,93 |
| *UV* | 4.197 | 14 | 0,33 |
| *UVA* | 11.017 | 16 | 0,15 |
| *UPV/EHU* | 2.192 | 288 | 13,14 |
| *UIA* | 3.283 | 149 | 4,54 |
| *UJI* | 290 | 217 | 74,83 |
| *UPV* | 1.334 | 515 | 38,61 |
| *URV* | 3.638 | 370 | 10,17 |

Fuente: Base de datos ISBN (19-02-2020).
*http://www.mcu.es/webISBN*
Se descarta la editorial de la *UCLM* porque no asigna ISBN a sus publicaciones en abierto menor proporción, la *Rovira i Virgili*), la presencia de los libros en abierto en el catálogo de las editoriales universitarias sigue siendo simbólica. La media total se sitúa en 139 documentos por universidad. Este valor equivale al 4,9% de la producción total editorial universitaria ya que la media de ISBNs registrados es de 2.860 publicaciones. Esto muestra un tímido avance a pesar de las políticas y recomendaciones en torno a la ciencia abierta impulsadas por la Unión Europea, quedando muy lejos del 15% de media que ofrecían los datos de *Dimensions* (tabla 1). Otro indicio revelador del bajo interés de las editoriales universitarias españolas por el libro en abierto es que, de manera generalizada, no se especifica si dichas publicaciones responden o no a resultados de investigación (a pesar de que, legalmente, tendría que aparecer esa información si ha habido alguna subvención oficial para financiar un proyecto).

## 5. Hacia unos libros académicos más abiertos en las universidades españolas

Antes de presentar las conclusiones de este trabajo, y de forma esquemática, proponemos reflexionar sobre la necesaria adopción de una serie de acciones o criterios a llevar a cabo por parte de las editoriales universitarias en sus procesos de edición y publicación de libros en abierto. Estas recomendaciones están centradas en las fases de edición, descripción, publicación, difusión y medición, tal como muestra el gráfico 3.

En la fase de edición es necesario definir no sólo el estilo de redacción académico-científico, procurando una estructura discursiva ágil y estandarizada, también hay que establecer (por parte de la editorial) el procedimiento de revisión por pares que debería ser análogo al que siguen las revistas científicas en cuanto a rigor y calidad. Es necesario materializar los acuerdos o contratos de edición bajo licencias de reutilización del contenido como ocurre en los contratos de edición de publicaciones comerciales. Para ello, es recomendable disponer de modelos de acuerdo o contratos necesarios supervisados por los servicios jurídicos de las universidades.

La fase de descripción lleva implícita un aumento de tareas documentales vitales para el resto de etapas de la publicación en abierto, tales como el establecimiento de estructuras y formatos de metadatos como ya se hace en las revistas académicas, además de la adopción y obligatoriedad del uso de identificadores persistentes, tanto para los autores/investigadores (orcid) como para las publicaciones digitales (DOI), garantizando así el control de autoridades y la localización permanente de la publicación. En esta fase será necesaria la extracción de las referencias bibliográficas citadas en las monografías para su visibilidad y cómputo en las distintas métricas bibliográficas. Este aspecto es vital para valorar adecuadamente el impacto científico real en disciplinas que tradicionalmente quedan relegadas en los rankings de estas métricas bibliográficas, muchas de ellas pertenecientes a las Ciencias Sociales y, fundamentalmente, casi todas las Humanidades.

En la fase de difusión es importante que las editoriales y las bibliotecas universitarias registren y revisen los datos de sus colecciones de monografías abiertas en cualquier fuente de información bibliográfica que fomente la visibilidad y el acceso a este tipo de publicación a nivel internacional (como *DOAB* u *Oapen*), además de compartir sus contenidos en redes sociales científicas y propiciar el seguimiento de los indicadores altmétricos (**Hammarfelt**, 2014; **Torres-Salinas**; **Robinson-García**; **Gorraiz**, 2017; **Torres-Salinas**; **Gorraiz**; **Robinson-García**, 2018). También deben preocuparse por la

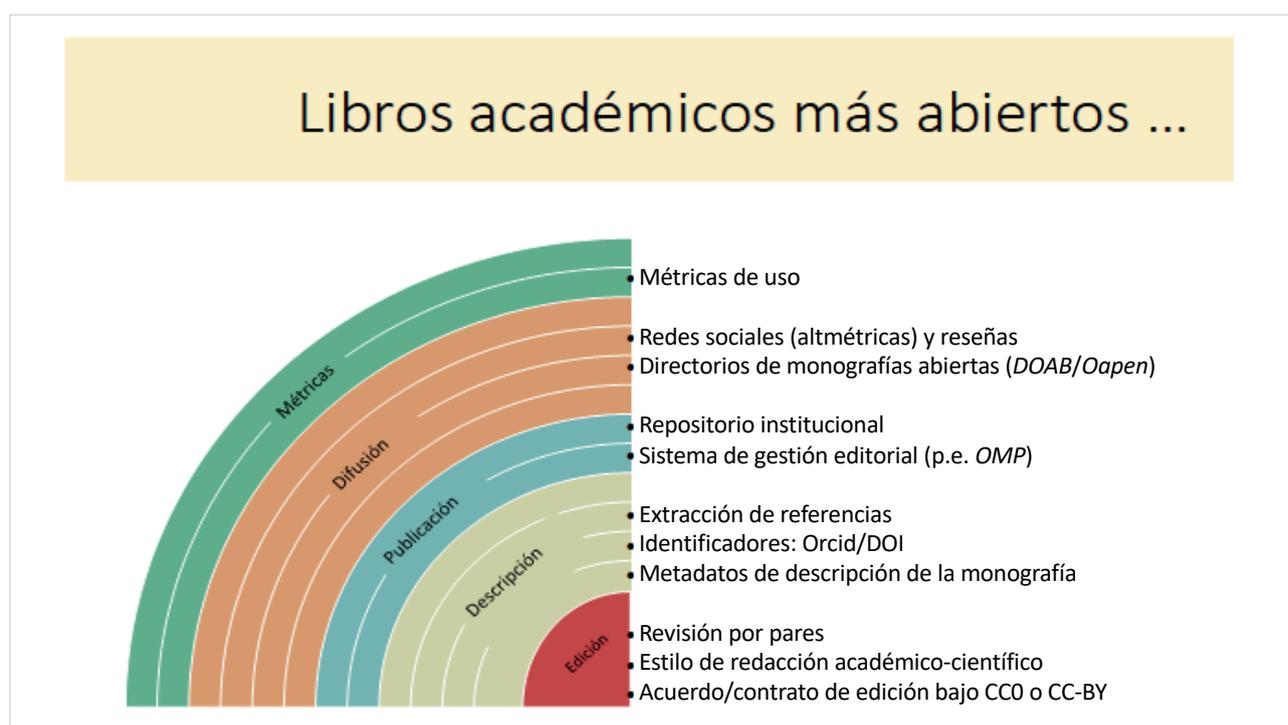

Gráfico 3. Acciones a llevar a cabo por las editoriales universitarias para la gestión y difusión de sus monografías en abierto. Adaptación de **Kramer** y **Bosman** (2018).

búsqueda y seguimiento de las reseñas de sus libros en abierto que seguirán proporcionando, al igual que ocurre en el sector editorial privado, una valoración cualitativa de la obra abierta (**Gorraiz**; **Gumpenberger**; **Purnell**, 2014). También se deberán contemplar, en el conjunto de métricas en torno a la monografía abierta, las estadísticas de uso no sólo de las estadísticas de las editoriales o los repositorios institucionales, sino también de otras fuentes de difusión de las publicaciones en abierto, como repositorios especializados o las redes sociales académicas.

Todos estos aspectos, si se introdujeran en la gestión y difusión de los libros en abierto, supondrían un replanteamiento generalizado en los servicios editoriales universitarios, que requerirían de apoyo institucional y materia (personal técnico e infraestructura tecnológica, principalmente).

## 6. Conclusiones

A modo de conclusión general, la publicación académica en abierto (más concretamente el libro académico) sigue siendo una anécdota en la producción editorial de las universidades públicas españolas, a pesar de los requerimientos, declaraciones, políticas y mandatos de los organismos de financiación. Una de las causas de esta situación es la escasa concreción e implementación de políticas de acceso abierto en el seno de las universidades (muchas de ellas se limitan a suscribir declaraciones de apoyo, pero no aplican normas ni procedimientos) y los cambios sobre el tradicional modelo de negocio y gestión editorial que puede llegar a provocar el acceso abierto de los libros académicos en el sector editorial universitario. La solución a este problema pasa por establecer estrategias y procedimientos de gestión, depósito y difusión de su producción editorial en abierto, alineando servicios bibliotecarios y editoriales en aras de una gestión rigurosa, sostenible y rentable que apueste por la visualización global de su producción académico-científica, sin renunciar a la gestión editorial tradicional que, en consecuencia, deberá someterse a un replanteamiento de su finalidad.

Los procesos editoriales de los libros académicos en abierto, en general, deben revisarse para adaptarse a las nuevas exigencias bibliográficas y bibliométricas, del mismo modo que ha ocurrido con las revistas científicas en abierto, utilizando lenguajes y estructuras de metadatos y empleando identificadores persistentes para la automatización del intercambio de datos bibliográficos que mejoren la visibilidad, transmisión de conocimiento y citación de las obras monográficas editadas, así como mejorar la imagen de transparencia en la propia gestión editorial universitaria.

Del mismo modo que las revistas científicas en abierto editadas por las universidades están en plena adaptación a las exigencias de la ciencia abierta, es preciso extender esta renovación al resto de publicaciones en abierto, empezando por el libro académico dotándolo asíde la calidad y visibilidad que merece en el contexto de difusión científica actual.

Por último, las editoriales universitarias son servicios públicos para la difusión de los resultados de la actividad investigadora. Por esta razón, deben también adaptar sus protocolos para que la publicación en abierto les acerque más a lo estipulado en el marco legal de la Ley de la Ciencia. Los títulos actualmente publicados no parecen estar muy alineados con esa estrategia.

## 7. Referencias

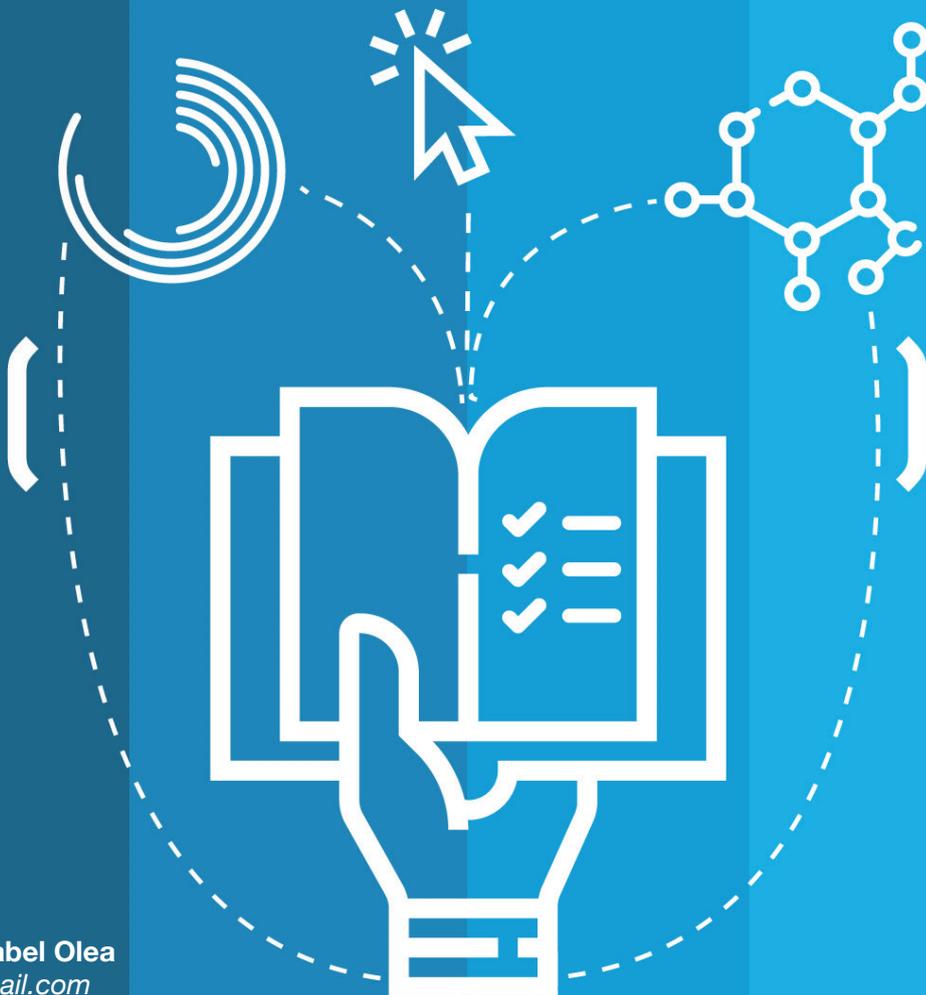

Contacto: **Isabel Olea**
*epi.iolea@gmail.com*

*http://www.elprofesionaldelainformacion.com/manual-revistas.html*

## Manual SCImago de revistas científicas. Creación, gestión y publicación
**Tomàs Baiget**